\documentclass[aps,prl,twocolumn,superscriptaddress,showpacs,floatfix,amsmath,amssymb,nofootinbib,preprintnumbers]{revtex4-1}

\usepackage{graphicx}

\usepackage{epstopdf}
\usepackage{color}

\newcommand{\be}{\begin{equation}}
\newcommand{\ee}{\end{equation}}
\newcommand{\ba}{\begin{eqnarray}}
\newcommand{\ea}{\end{eqnarray}}

\newcommand{\beq}{\begin{equation}}
\newcommand{\eeq}{\end{equation}}
\newcommand{\beqa}{\begin{eqnarray}}
\newcommand{\eeqa}{\end{eqnarray}}

\begin{document}
\title{Kerr-AdS analogue of triple point and solid/liquid/gas phase transition}

\author{Natacha Altamirano}
\email{naltamirano@famaf.unc.edu.ar}
\affiliation{Perimeter Institute, 31 Caroline St. N., Waterloo,
Ontario, N2L 2Y5, Canada}
\affiliation{Facultad de Matem\'atica, Astronom\'ia y F\'isica, FaMAF, Universidad Nacional de C\'ordoba,
Instituto de F\'isica Enrique Gaviola, IFEG, CONICET,
Ciudad Universitaria (5000) C\'ordoba, Argentina}
\author{David Kubiz\v n\'ak}
\email{dkubiznak@perimeterinstitute.ca}
\affiliation{Perimeter Institute, 31 Caroline St. N., Waterloo,
Ontario, N2L 2Y5, Canada}
\author{Robert B. Mann}
\email{rbmann@sciborg.uwaterloo.ca}
\affiliation{Department of Physics and Astronomy, University of Waterloo,
Waterloo, Ontario, Canada, N2L 3G1}
\author{Zeinab Sherkatghanad}
\email{zsherkat@uwaterloo.ca}
\affiliation{Perimeter Institute, 31 Caroline St. N., Waterloo,
Ontario, N2L 2Y5, Canada}
\affiliation{Department of Physics and Astronomy, University of Waterloo,
Waterloo, Ontario, Canada, N2L 3G1}
\affiliation{Department of Physics, Isfahan University of Technology, Isfahan, 84156-83111, Iran}

\date{January 25, 2014}  

\begin{abstract}
We study 
the thermodynamic behavior of multi-spinning $d=6$ Kerr-anti de Sitter  black holes in the canonical ensemble of fixed angular momenta  
$J_1$ and $J_2$. 
We find, dependent on the ratio
$q=J_2/J_1$, qualitatively different interesting phenomena known from the `every day thermodynamics' of simple substances.   For $q=0$   the system exhibits recently observed reentrant large/small/large black hole phase transitions, but for $0<q\ll 1$ we find an analogue of a `solid/liquid' phase transition.  Furthermore,  for $q\in (0.00905, 0.0985)$ the system displays the presence of a large/intermediate/small black hole phase transition with  two critical  and one triple (or tricritical) points. This  behavior is reminiscent of the solid/liquid/gas phase transition except that the
 coexistence line of small and intermediate black holes does not continue for an arbitrary value of pressure (similar to the solid/liquid coexistence line) but rather terminates at one of the critical points. Finally, for $q>0.0985$ we observe the `standard liquid/gas behavior' of the Van der Waals fluid.  
 \end{abstract}


\maketitle

\section{Introduction}
Asymptotically AdS black hole (BH) spacetimes demonstrate various phase transitions,
which, due to the AdS/CFT correspondence 
offer a dual interpretation in the boundary conformal field theory. Particularly interesting 
is the Hawking--Page transition \cite{HawkingPage:1983} which corresponds to the confinement/deconfinment phase transition in the dual quark
gluon plasma \cite{Witten:1998b}. For charged or rotating black holes in four dimensions one observes a small/large black hole first order phase transition reminiscent of the liquid/gas transition of the Van der Waals fluid \cite{ChamblinEtal:1999a,ChamblinEtal:1999b,CveticGubser:1999a, CaldarelliEtal:2000, TsaiEtal:2012, Dolan:2011a, KubiznakMann:2012, Hristov:2013sya, Johnson:2013}.

Interesting new phenomena appear in higher dimensions.  Recently it has been shown  \cite{Altamirano:2013ane} that in all $d\geq 6$ dimensions the single spinning 
vacuum Kerr-AdS black holes demonstrate the peculiar behavior of large/small/large 
black hole transitions reminiscent of   {\em reentrant phase transitions} observed for multicomponent fluid systems, gels, ferroelectrics, liquid crystals, and binary gases, e.g., \cite{NarayananKumar:1994}.

In this letter we find a more elaborate phase structure for multiply rotating Kerr-AdS black holes.  Specifically, for $d=6$ we find, dependent on
the  ratio $q=J_2/J_1$ of the two angular momenta, the existence of a triple point along with other qualitatively different interesting phenomena known from the `every day thermodynamics'.   i) For $q=0$ we recover (for a certain range of temperature) the aforementioned reentrant phase transition \cite{Altamirano:2013ane}. ii)  For $q\ll 1$ we observe an analogue of a solid/liquid phase transition.
iii) For $q\in (0.00905, 0.0985)$,   the system displays a small/intermediate/large black hole phase transition with one {\em triple} (or tricritical) and two critical points, reminiscent of the solid/liquid/gas phase transition. iv) Finally, for $q>0.0985$ the `standard' liquid/gas behavior of the Van der Waals fluid  \cite{ChamblinEtal:1999a,ChamblinEtal:1999b,CveticGubser:1999a, CaldarelliEtal:2000,
TsaiEtal:2012, Dolan:2011a, KubiznakMann:2012, Hristov:2013sya, Johnson:2013, 
NiuEtal:2011, Dolan:2012, GunasekaranEtal:2012, Cai:2013qga,Poshteh:2013pba} is observed.

\section{Black hole spacetimes}
To treat  even ($\epsilon=1)$  odd ($\epsilon=0)$ spacetime dimensionality $d$ simultaneously, we parametrize 
$d=2N + 1 + \epsilon $.
General Kerr-AdS black holes  admit $N$ independent angular momenta $J_i$, described by $N$ rotation parameters $a_i$; the metric is
\cite{GibbonsEtal:2004, GibbonsEtal:2005}:
\ba \label{metric}
ds^2&=&-W\Bigl(1+\frac{r^2}{l^2}\Bigr)d\tau ^2+\frac{2m}{U} \Bigl(W d\tau -\sum_{i=1}^{N} \frac{a_i \mu_i ^2 d\varphi _i}{\Xi _i}\Bigr)^2\nonumber\\
&+&\sum_{i=1}^{N} \frac{r^2+a_i^2}{\Xi _i} \mu_i ^2 d\varphi _i^2+\frac{U dr^2}{F-2m}+\sum_{i=1}^{N+\epsilon}\frac{r^2+a_i ^2}{\Xi _i} d\mu _i ^2 \nonumber\\
&-&\frac{l^{-2}}{W (1+r^2/l^{2})}\Bigl(\sum_{i=1}^{N+\epsilon}\frac{r^2+a_i ^2}{\Xi _i} \mu_i d\mu_i\Bigr)^2\,.
\ea
Here, the azimuthal coordinates $\mu_i$ are constrained to obey $\sum_{i=1}^{N+\epsilon}\mu_i^2=1$, in even dimensions. We have set for convenience $a_{N+1}=0$, and 
\ba\label{metrcifunctions}
W&=&\sum_{i=1}^{N+\epsilon}\frac{\mu _i^2}{\Xi _i}\,,\quad U=r^\epsilon \sum_{i=1}^{N+\epsilon} \frac{\mu _i^2}{r^2+a_i^2} \prod _j ^N (r^2+a_j^2)\,,\nonumber\\
F&=&r^ {\epsilon -2} \Bigl(1+\frac{r^2}{l^2}\Bigr) \prod_{i=1}^N (r^2+a_i^2)\,,\quad \Xi_i=1-\frac{a_i^2}{l^2}\,.\quad
\end{eqnarray}
The metric obeys the Einstein equations with cosmological constant:
$R_{ab}=-\frac{d-1}{l^2} g_{ab}$.

\begin{figure}
\vspace{-0.7cm}
\begin{center}
\includegraphics[width=0.35\textheight,height=0.43\textwidth]{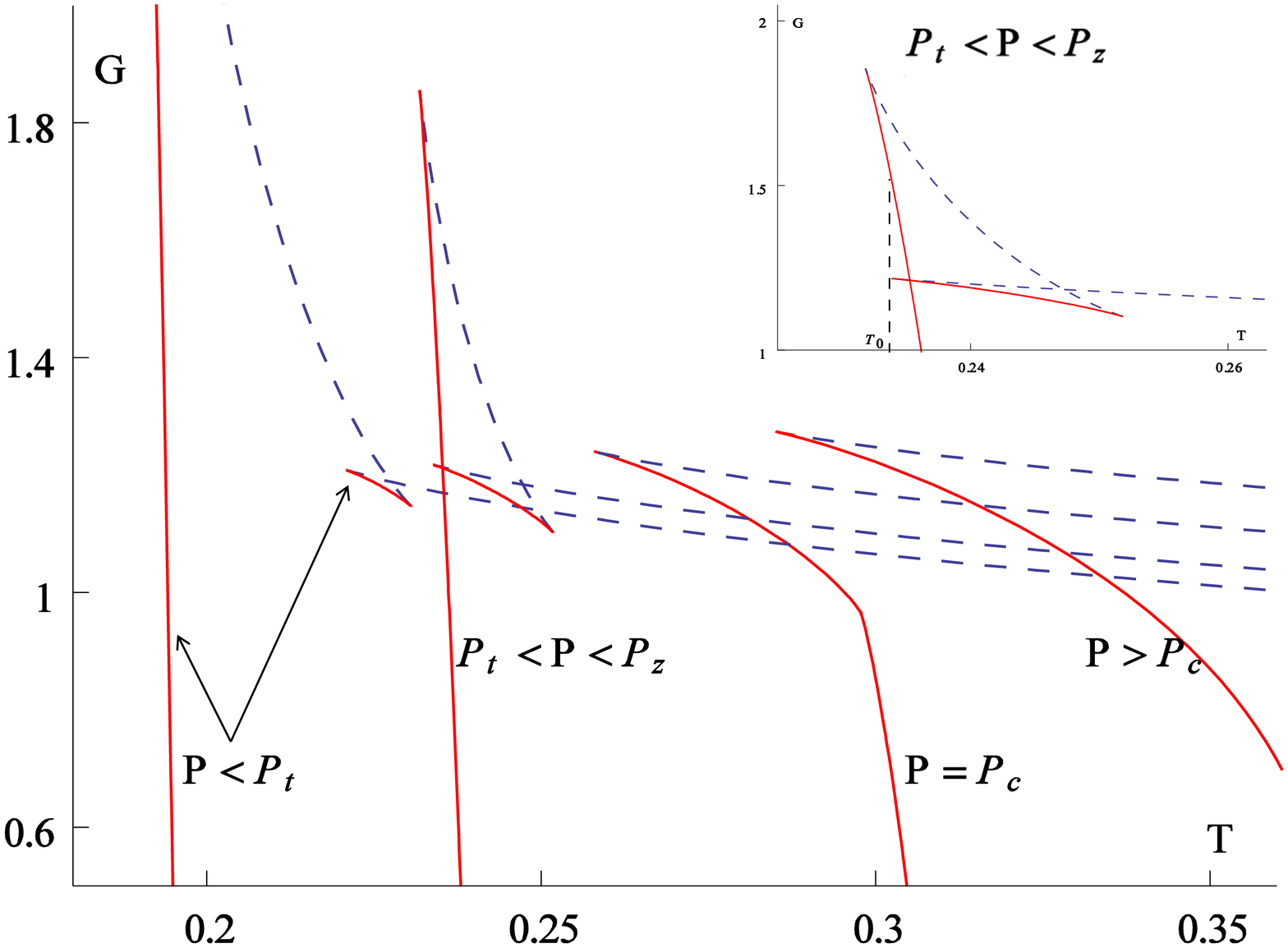}
\caption{{\bf Gibbs free energy for $q=0$.}
Pressures increase  from left to right and solid-red/dashed-blue lines correspond to
$C_P$ positive/negative respectively; at their  joins   $C_P$ diverges. Moving from right to left, `near vertical' solid red curves have an increasingly large negative slope. As $T\to \infty$ they asymptote to $-\infty$; for sufficiently small $T$, each curve joins its blue-dashed counterpart at some
large positive value of $G$, the limit of zero temperature is attained in the asymptotically flat case $P\to 0$. 
 As with Schwarzschild-AdS black holes, for $P\geq P_c$, the (lower) large BH branch is thermodynamically stable  whereas the upper branch is unstable.  For $P=P_c$ we observe critical behavior. For  $P\approx 0.0564\in(P_t,P_z)$ we observe a
``zeroth-order phase transition'': a discontinuity in the global minimum of $G$ at $T=T_0\approx 0.2339\in(T_t,T_z)$ (denoted by the vertical line in the inset)  signifying the onset of an reentrant phase transition. For $P<P_t$ only one branch of stable large BHs exists.  
} \label{fig:1}
\end{center}
\end{figure} 
\begin{figure}
\vspace{-0.7cm}
\begin{center}
\includegraphics[width=0.47\textwidth,height=0.32\textheight]{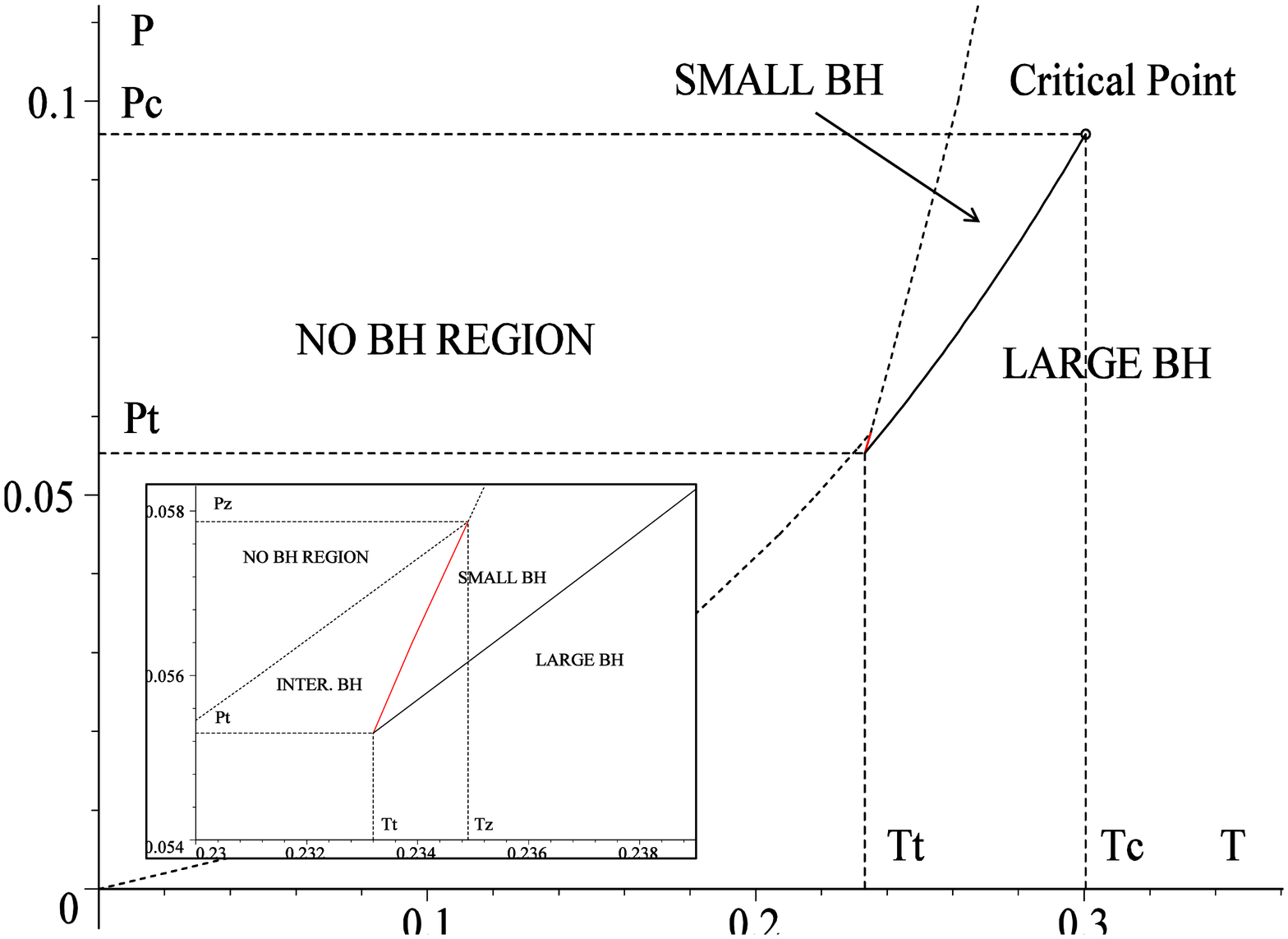}
\caption{{\bf $P-T$ diagram for $q=0$.}
The coexistence line of the first order phase transition between small and large black holes is depicted by a thick black solid line. It initiates from the critical point $(P_c, T_c)$ and terminates at $(P_t, T_t)$. The red solid line (inset) indicates the `coexistence line' of small and intermediate black holes, separated by a finite gap in $G$, indicating the 
reentrant phase transition.  It commences from  $(T_z, P_z)$ and terminates at $(P_t, T_t$).
 The ``No BH region" is to the left of the dashed oblique curve, containing the $(T_z, P_z)$ point.  Below $T_t$, the lower dashed curve
 terminates at the origin $P=T=0$.
}  
\label{fig:2}
\end{center}
\end{figure} 

The mass $M$, the angular momenta $J_i$, and the angular velocities of the horizon $\Omega_i$ are \cite{Gibbons:2004ai}
\ba \label{TD}
M&=&\frac{m \Omega _{d-2}}{4\pi (\prod_j \Xi_j)}(\sum_{i=1}^{N}{\frac{1}{\Xi_i}-\frac{1-\epsilon }{2}})\,,\nonumber\\
J_i&=&\frac{a_i m \Omega _{d-2}}{4\pi \Xi_i (\prod_j \Xi_j)}\,,\quad \Omega_i=\frac{a_i (1+\frac{r_+^2}{l^2})}{r_+^2+a_i^2}\,,
\ea
while the temperature $T$ and the entropy $S$ read
\ba\label{TS}
T&=&\frac{1}{2\pi }\Bigr[r_+\Bigl(\frac{r_+^2}{l^2}+1\Bigr)
\sum_{i=1}^{N} \frac{1}{a_i^2+r_+^2}-\frac{1}{r_+}
\Bigl(\frac{1}{2}-\frac{r_+^2}{2l^2}\Bigr)^{\!\epsilon}\,\Bigr]\,,\nonumber\\
S&=&\frac{\Omega _{d-2}}{4 r_+^{1-\epsilon}}\prod_{i=1}^N 
\frac{a_i^2+r_+^2}{\Xi_i}\,,
\ea
with the horizon radius $r_+$ being determined as the largest root of $F-2m=0$.

To discuss the phase diagram of these spacetimes we shall re-interpret the negative cosmological constant as positive pressure  \cite{CaldarelliEtal:2000,Padmanabhan:2002sha, KastorEtal:2009,CveticEtal:2010, Dolan:2011a, KubiznakMann:2012, Dolan:2012, Altamirano:2013ane, GunasekaranEtal:2012, Cai:2013qga}
\be\label{P}
P=-\frac{\Lambda}{8 \pi} = \frac{(d-1) (d-2)}{16 \pi l^2}\,.
\ee
In the canonical (fixed $J_i$) ensemble, the information about the equilibrium thermodynamics is captured by the Gibbs free energy
\be\label{G}
G=M-TS=G(P,T,J_1,\dots,J_N)\,.
\ee 
Namely, we assume the thermodynamic postulate that for a fixed $(T,P, J_i)$ the state of a system corresponds to the {\em global minimum} of $G$. 
We also accept the positivity of the specific heat at constant $P$ and $J_i$ 
\be
C_P\equiv C_{P,J_1,\dots,J_N}=T\left(\frac{\partial S}{\partial T}\right)_{P,J_1,\dots,J_N}\,,
\ee 
as a local criterion of thermodynamic stability.



Specifying to $d=6$ (for which $N=2$) , we consider the Gibbs free energy $G=G(P,T,J_1,J_2)$, obtained from the expression \eqref{G} by eliminating parameters $(l, r_+, a_1, a_2)$ in favor of $(P, T,J_1,J_2)$ using Eqs.~\eqref{TD}, \eqref{TS} and \eqref{P}.  An analytic solution is not possible
since 
higher-order polynomials are encountered and so we proceed numerically: for a given $P, r_+, J_1$ and $J_2$, we solve the second equation in \eqref{TD} for $a_1$ and $a_2$  and thence calculate the values of $T$ and $G$ using \eqref{TS} and \eqref{G}, yielding a $G-T$ diagram.
Once the behavior of $G$ is known, we compute the corresponding phase diagram,  coexistence lines, and critical points in the $P-T$ plane. We display our results in Figs. \ref{fig:1}--\ref{fig:7}. Since the qualitative behavior of the system depends only on  $q=J_2/J_1$ we set everywhere $J_1=1$. 


\section{Reentrant phase transition}
The $q=0$ results (recently considered in \cite{Altamirano:2013ane}) are displayed in Fig.~\ref{fig:1}. For each isobar, the horizon radius of the black hole increases as we move along the line from upper right to lower left. For small pressures $(P<P_t\approx 0.0553)$, there is only one phase of stable large black holes, the corresponding Gibbs free energy being the left-most curve in  Fig.~\ref{fig:1}. 

The inset (a magnification of the second left-most curve) illustrates the reentrant phase transition: for the pressure/temperature ranges $P\in(P_t,P_z)\approx (0.0553,0.0579)$ and $T\in (T_t, T_z)\approx (0.2332,0.2349)$, the global minimum of the Gibbs free energy experiences a finite jump (in the inset at $T=T_0$), a `zeroth-order' phase transition, whose physics  remains to be understood.  
At a slightly higher temperature there is a  standard first order phase transition (a jump in the first derivative of $G$). 
As the temperature decreases from right to left,
the system follows the lower solid red curve (large BH) until it joins the middle solid red curve (small BH),  corresponding to a first order small/large black hole phase transition. 
As $T$  decreases further, the system follows this middle red curve (small BH) until  $T= T_0\in (T_t, T_z)$, where $G$ has a discontinuity at its global minimum, at which point it jumps to the uppermost red line where black holes are again larger for all smaller $T$ --- this corresponds to the zeroth order phase transition between small and intermediate black holes. In other words, a continuous decrease of one thermodynamic state variable, temperature, induces a phase change from large to small and back to large black holes. This is the reentrant phase transition observed in \cite{Altamirano:2013ane}.

For pressures $P>P_z$ the jump in the global minimum of $G$ disappears while the first order small/large black hole phase transition still persists. This behavior terminates at the critical point,  corresponding to the third curve and characterized by 
$(T_c, P_c)\approx (0.3004,0.0958)$, above which only one phase of black holes exists (the right-most curve).

The overall situation is summarized in the $P-T$ phase diagram (Fig.~\ref{fig:2}). Since for any given pressure $P$ there is a minimum temperature $T_{\tiny{min}}(P)$ for which any black hole can exist, there is always a `no black hole region'  (not displayed in \cite{Altamirano:2013ane}), which starts from zero at $P\to 0$ and `grows bigger' as $P$ increases; this occurs to the left of the dashed oblique curve and contains the $(T_z, P_z)$ point.

\section{Solid/liquid analogue}

\begin{figure}
\vspace{-0.7cm}
\begin{center}
\rotatebox{-90}{
\includegraphics[width=0.37\textwidth , height=0.34\textheight]{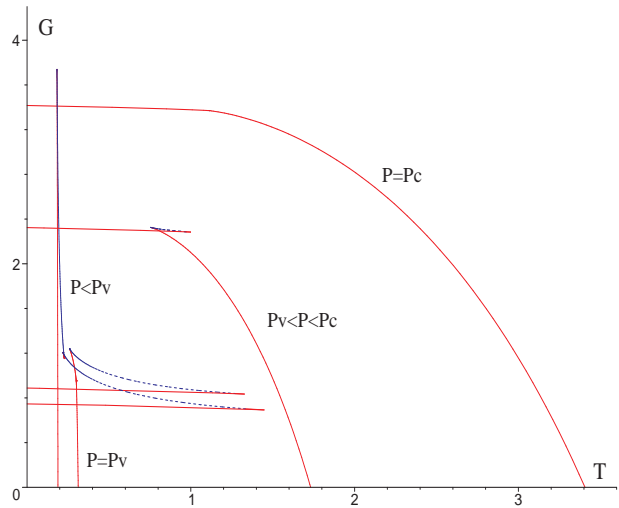}
}
\caption{{\bf Gibbs free energy for $q=0.005$} is displayed for  decreasing pressures (from top to bottom). 
The horizon radius $r_+$ increases from left to right. The uppermost isobar corresponds to $P=P_{c}=4.051$; for higher pressures only one branch of (stable) black holes with positive $C_P$ exists. The second uppermost isobar displays the swallowtail behavior and implies the existence of a first order phase transition. For $P=P_v\approx 0.0958$ another critical point emerges but occurs for a branch that does not minimize $G$ globally. Consequently, out of the two swallowtails for $P<P_v$ only one occurs in the branch globally minimizing $G$ and describes a `physical' first order phase transition. 
} \label{fig:3}
\end{center}
\end{figure}

For $0<q\ll 1$ the situation completely changes: the unstable branch of tiny hot black holes on the right of the $G-T$ diagram in Fig.~\ref{fig:1} disappears and a new branch of (locally) stable tiny cold black holes appears to the left. The $q=0$ `no black hole region' is eliminated and the situation is very similar to what happens when a small charge is added to a Schwarzschild black hole \cite{ChamblinEtal:1999a}. The zeroth-order phase transition  is `replaced' by a `solid/liquid'-like phase transition of small to large black holes.

The behavior of $G$ for $q<q_1\approx 0.00905$ is displayed in Fig.~\ref{fig:3}. 
Although in this range of angular momenta $G$ admits two critical points, only one of them occurs for stable black holes that minimize $G$ globally. Consequently we observe one phase transition between small and large black holes. The corresponding coexistence line in the $P-T$ diagram is similar that of non-rotating charged black holes \cite{KubiznakMann:2012}, monotonically increasing and terminating at a critical point characterized by $P_c$ and $T_c$.  These quantities both increase as $q$ decreases; in the limit $q\to 0$ we find an `infinite' coexistence line, similar to what happens for a solid/liquid phase transition.   

\section{Triple point and solid/liquid/gas analogue}
\begin{figure}
\vspace{-0.7cm}
\rotatebox{-90}{
\includegraphics[width=0.37\textwidth , height=0.34\textheight]{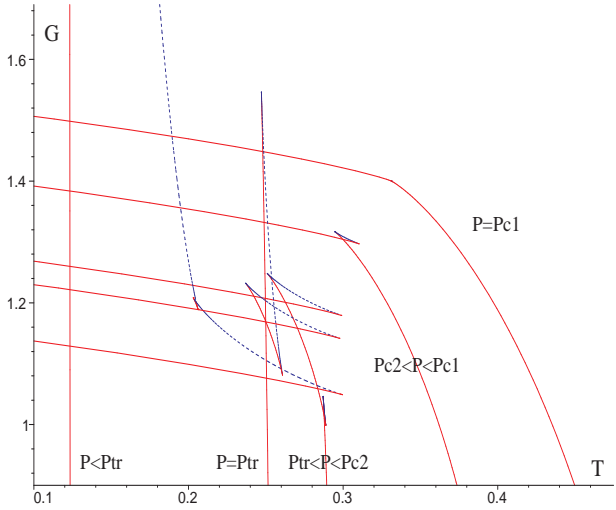}
}
\caption{ {\bf Gibbs free energy for $q=0.05$,} displayed for   various pressures (from top to bottom) $P=0.259, 0.170, 0.087, 0.0641, 0.015$. The horizon radius $r_+$ increases from left to right. The uppermost isobar corresponds to $P=P_{c_1}=0.259$; for higher pressures only one branch of stable black hole with positive $C_P$ exists. The second uppermost isobar displays the swallowtail behavior, implying a first order phase transition. 
The third isobar corresponds to $P_{c_2}=0.0956<P<P_{c_1}$ for which we have `two swallowtails'. For such pressures there are two first order phase transitions. The fourth isobar displays the tricritical pressure $P_{tr}=0.087$ where the two swallowtails `merge' and the triple point occurs. Finally the lower-most isobar corresponds to $P<P_{tr}$.  
} \label{fig:5}
\end{figure} 
\begin{figure}
\vspace{-0.5cm}
\rotatebox{-90}{
\includegraphics[width=0.37\textwidth , height=0.34\textheight]{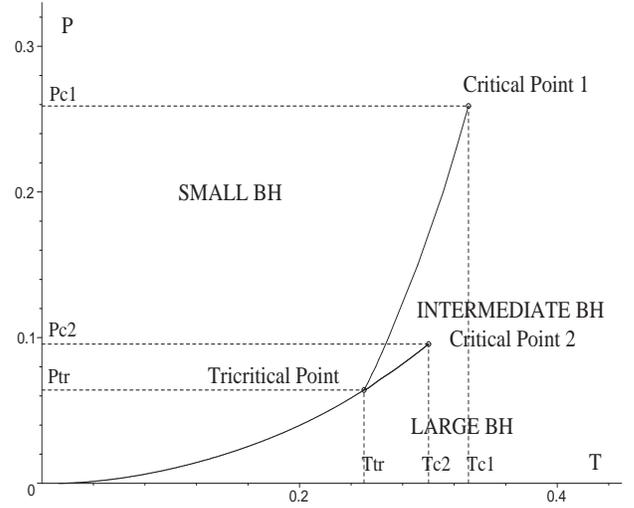}
}
\caption{ {\bf  $P-T$ diagram for $q=0.05$.} The diagram is analogous to the solid/liquid/gas phase diagram. 
Note however that there are two critical points. That is, the solid-liquid coexistence line does not extend to infinity but rather terminates, similar to the ``liquid/gas'' coexistence line, in a critical point. }  
\label{fig:6}
\vspace{-0.1cm}
\end{figure} 
At $q=q_1$ a new phenomenon occurs: a triple point and a second critical point emerge from the coexistence line at $P_{tr}=P_{c_2}\approx 0.09577$ and $T_{tr}=T_{c_2}\approx 0.30039$. As $q$ increases, the triple point moves away from the second critical point (the values of $T_t$ and $P_t$ decrease). A small/intermediate/large black hole phase transition may occur and the situation resembles that of the solid/liquid/gas phase transition. 
At the same time the first critical point moves towards the triple point ($T_{c_1}$ and $P_{c_1}$ decrease). At $q=0.08121$ both critical points occur at the same pressure 
$P_{c_1}=P_{c2}\approx 0.0953$, whereas $T_{c_1}\approx 0.2486<T_{c_2}\approx 0.2997$. Increasing $q$ even further, the first critical point moves closer and closer to the triple point and finally for $q=q_2\approx 0.0985$ the two merge at  $P_{tr}=P_{c_1}\approx 0.049$. Above $q_2$ only the second critical point remains.  
 
The Gibbs free energy for $q=0.05$ is displayed in Fig.~\ref{fig:5} and the corresponding $P-T$ diagram  in Fig.~\ref{fig:6}. 
Note that for this value of $q$, the critical pressure $P_{c_1}$ is bigger than $P_{c_2}$ and the  solid/liquid/gas behavior is present.

\section{Van der Waals behaviour}
\begin{figure}
\vspace{-0.3cm}
\begin{center}
\includegraphics[width=0.54\textwidth,height=0.29\textheight]{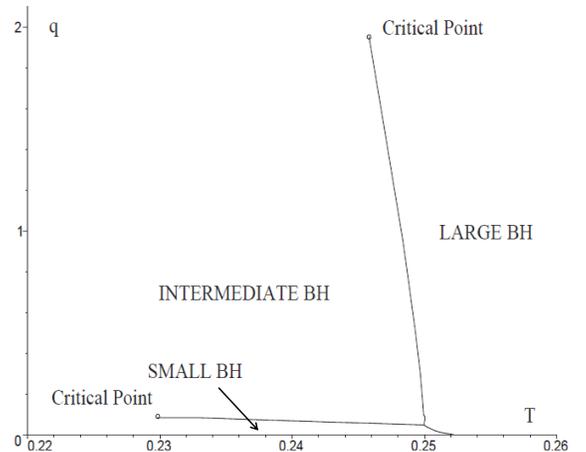}
\caption{{\bf Triple point in $q-T$ plane.}
A `solid/liquid/gas' phase structure appears for fixed $\Lambda$, here given by 
$l\approx 2.491$. We observe three coexistence lines corresponding to the associated first-order phase transitions. Two of them terminate at corresponding critical points. All three then merge in a triple point. 
} \label{fig:7}
\end{center}
\vspace{-0.7cm}
\end{figure}
For $q>q_2$ only one small/large black hole coexistence line exists and terminates at a corresponding 
critical point. We observe an analogue of  a Van der Waals  `liquid/gas' phase transition.

In fact, the situation is a little more subtle. For $q_2<q<q_3\approx 0.1274$, similar to the `solid/liquid analogue', the second critical point still exists, but occurs for the locally unstable branch of black holes with negative heat capacity that does not globally minimize the Gibbs free energy. Hence, only one phase transition is physical. For $q>q_3$ this second critical point completely disappears. 
In both cases we observe a `standard liquid/gas' Van der Waals phase transition  
\cite{ChamblinEtal:1999a,ChamblinEtal:1999b,CveticGubser:1999a, CaldarelliEtal:2000,
TsaiEtal:2012, Dolan:2011a, KubiznakMann:2012, Hristov:2013sya, Johnson:2013, 
NiuEtal:2011, Dolan:2012, GunasekaranEtal:2012, Cai:2013qga,Poshteh:2013pba}.  Qualitatively similar behavior persists for all-spin-equal, $a_1=a_2=\dots=a_N$, rotating black holes in any odd dimension $d\geq 5$.

\section{Discussion}

We have shown that multiply-rotating black holes exhibit a rich set of interesting thermodynamic phenomena known from the `every day thermodynamics' of simple substances. Specifying to six dimensions, we have demonstrated that reentrant phase transitions, triple points, multiple first-order transitions, solid/liquid/gas phase transitions, and Van der Walls `liquid/gas' phase transitions can all occur depending on the ratio of the two angular momenta. Note that neither the existence of the reentrant phase transition, nor of the triple point depends on a variable $P\sim\Lambda$.  For any fixed $P$ within the allowed range, these phenomena will
take place; in the AdS/CFT context  there will be a corresponding reentrant phase transition within the allowed range of $N$ in the dual $SU(N)$ gauge theory.   We illustrate triple point behavior in a $q-T$ diagram for fixed $P$  in Fig. \ref{fig:7}. 
 We have evidence that such phenomena will take place in dimensions $d> 6$, though the details remain to be worked out.

\section{Acknowledgments} 
This research was supported in part by Perimeter Institute for Theoretical Physics and by the Natural Sciences and Engineering Research Council of Canada. Research at Perimeter Institute is supported by the Government of Canada through Industry Canada and by the Province of Ontario through the Ministry of Research and Innovation.


\providecommand{\href}[2]{#2}\begingroup\raggedright\endgroup

\end{document}